\documentclass[aps,prb,twocolumn,showpacs,twoside]{revtex4}
\usepackage{graphicx,psfrag,amssymb,amsmath,amsbsy}
\usepackage[dvips]{color}%\usepackage{marvosym} % для \Stopsign
\newcommand{\pf}{p_\mathrm F}
\newcommand{\be}{\begin{equation}}\newcommand{\ee}{\end{equation}}
\newcommand{\bes}{\begin{equation*}}\newcommand{\ees}{\end{equation*}}
\newcommand{\bea}{\begin{eqnarray}} \newcommand{\eea}{\end{eqnarray}}
\newcommand{\beas}{\begin{eqnarray*}} \newcommand{\eeas}{\end{eqnarray*}}
\newcommand{\GR}{G_\mathrm R}\newcommand{\GA}{G_\mathrm A}

\newcommand{\GK}{G_\mathrm{K}}\newcommand{\GRA}{G_{\mathrm{R/A}}}

\newcommand{\Sp}{\mathop{\mathrm{Sp}}}
\newcommand{\ud}{\mathrm{d}}

\newcommand{\insfig}[2]{\begin{minipage}[c]{#1}\includegraphics[width=#1]{#2}\end{minipage}}
\newcommand{\insfigh}[3]{\begin{minipage}[c]{#1}\includegraphics[height=#2]{#3}\end{minipage}} 

\newcommand{\EF}{E_\mathrm F}
\newcommand{\vf}{v_\mathrm F}

\definecolor{mygrey}{gray}{0.85} % делаем фон слегка серым, чтобы при разглядывании файла в dvips меньше уставали глаза

\newcommand{\iqd}{\int\frac{\ud^2q}{(2\pi)^2}}

\newcommand{\ssh}[1]{{\sigma_{yx}^{z(#1)}}}\newcommand{\shs}{{\sigma_{yx}^z}}
\newcommand{\mybf}[1]{{\bf #1}}\renewcommand{\vec}{\mybf}\renewcommand{\Sp}{\mathop{\mathrm{Tr}}}
\begin{document}
\title{Spin-Hall conductivity due to Rashba spin-orbit interaction in disordered systems}
\author{Oleg Chalaev} \author{Daniel Loss}
\affiliation{Department of Physics and Astronomy, University of Basel, Klingelbergstrasse 82, Basel, CH-4056, Switzerland}
%\date{January 12, 2005}
\date{April 25, 2005}
%\date{\today}
\begin{abstract}
We consider the spin-Hall current in a disordered two-dimensional electron gas in the presence
of Rashba spin-orbit interaction. We derive
a generalized Kubo-Greenwood formula for the spin-Hall conductivity $\shs$
and evaluate it in an systematic way  using standard diagrammatic techniques for disordered systems. We find that in the
diffusive regime {both Boltzmann and the weak localization contributions
to  $\shs$ are of the same order and vanish in the zero frequency limit.}
We show that the uniform spin current is given by the total time derivative of the magnetization from
which we can conclude that the spin current vanishes exactly in the stationary limit.
This conclusion is valid for arbitrary spin-independent disorder, external electric field strength, and
also for interacting electrons.
\end{abstract}
 % <-- put spaces in this list
% PACS:
% Metals / spin polarized transport in, 72.25.Ba
% (Anderson) Localization / conductivity in metals and alloys, 72.15.Rn
% Disordered solids / localization in, 71.55.Jv
% Metals / transport processes in, 72.15.-v
%
% Electron scattering /  elastic scattering, 34.80.Bm
% Anderson localization /  disordered solids, 71.23.An
% Localization /  weak, 72.15.Rn, 73.20.Fz
%
% Disordered solids / electrical conductivity, 72.80.Ng
% Amorphous metals and alloys / electrical and thermal conductivity, 72.15.Cz
\maketitle
%%%%%%%%%%%%%%%%%%%%%%%%%%%%%%%%%%%%%%%%%%%%%%%%%%%%%%%%%%%%%%%%%%%%%%%%%%%%%%%%%%%%%%%%%%%%%%%%%%%%%%%%%%%%%%%%%%%%%%
\section{Introduction}
The idea of using the spin degrees of freedom of electrons in semiconductor systems instead of their charge  has
attracted wide interest in recent years \cite{wolf01:1488,prinz98:1660,awschalom02:xx,Zutic}.
Of particular interest are systems with spin-orbit interaction as this allows access to spin via charge
and vice versa. A number of effects in such systems have been studied
theoretically \cite{dyakonov71:467,hirsch99:1834,levitov85:133,edelstein:95,rashba03:241315R,%
rashba:0404723,murakami:0405003,ibmNew,Sinova,murakami03:1348,murakami:0405001,%
culcer:0309475,Sinitsin,DJ,MSH,SJD,Olya,Nomura04}, and experimentally
\cite{wolf01:1488,prinz98:1660,awschalom02:xx,Zutic,kato:0403407,ganichev:0403641},
and most recently by Kato {\it et al.} \cite{AwschScience04} who report the
first observation of spin Hall effect in GaAs and InGaAs samples of the extrinsic type \cite{dyakonov71:467,hirsch99:1834}.
In particular, there has been  considerable interest in Rashba spin-orbit interaction \cite{Rashba}
in two-dimensional electron systems since this type of spin-charge coupling can be controlled
by electrical gates.
Moreover, it has been pointed out by Sinova {\it et al.}  \cite{Sinova} that clean (without impurities) systems with
spin-orbit interaction carry spin currents and show a spin-Hall effect: applying an electric field $E_{x}$ in
$x$-direction generates a spin current in perpendicular $y$-direction.  In a number of subsequent papers the
corresponding spin-Hall conductivity $\shs$ has been calculated, and there seems to be agreement now that the spin-Hall
conductivity at zero frequency vanishes in an infinite system \cite{ibmNew,MSH,Olya,Sasha,RashbaSaysZero,RS0}.
%________________________________________________________________________________________
In this paper we present a systematic calculation of $\shs$ by using well-known perturbative techniques for disordered
systems \cite{AGD}. The systematic expansion is performed  in powers of the small parameter 
$1/\pf l$, with $\pf$ the Fermi momentum and $l$ the mean free path.
In addition to confirming the results of earlier work \cite{ibmNew,MSH,Olya,Sasha,RS0},
which were obtained in the Boltzmann (semiclassical) regime, we go beyond this limit
 and also include the weak localization correction diagrams. We show that this latter 
 contribution is of the same order in $1/\pf l$ and Rashba amplitude
as the Boltzmann value and vanishes as well at zero frequency. This demonstration
requires us to consider the various contributions to the 
Hikami box  in the weak localization contribution as well as the renormalization of the current vertices.

We then go on and show that the vanishing of the spin current is expected on rather general grounds and under general
conditions, such as arbitrary (spin-independent) disorder, electric field strength, and for interacting electrons.
Crucial for our argument are two observations. First, the uniform spin current is given by the total time derivative of
the magnetization\cite{SJD}, which, in turn, provides a straightforward interpretation of the spin current and its
observable effect in terms of magnetization and its change in time. Second, a system driven by an external force which is
constant in time normally reaches a steady state, implying then that the spin current vanishes under rather general
conditions.

\section{Model system}
We consider non-interacting electrons of mass $m$ and charge $e$ moving in a disordered two-dimensional system in the
presence of Rashba spin-orbit interaction with amplitude $\alpha$ \cite{Rashba,Sinova,DJ}. This system is described by the
Hamiltonian (we set $\hbar=1$)
\be\label{hamrash}
\hat H=\frac{\hat p^2}{2m}+\alpha(\hat\sigma^1\hat p_y-\hat\sigma^2\hat p_x)+U(\vec r),
\ee
where $\hat p$ denotes the momentum operator, and $\hat\sigma^{\mu}$ are the Pauli matrices. Further, $U(\vec r)$ is a
short-ranged disorder potential with the property that
$\overline {U(\vec r)U(\vec r\,')}=(m\tau)^{-1}\delta(\vec r-\vec r\,')$,
where the overbar indicates average over the disorder configuration,
%$\nu=m/2\pi$ is the density of states per spin,
and $\tau$ the mean free time between collisions.

The spin-orbit term in the Hamiltonian (\ref{hamrash}) changes the expression for the charge-current density operator (in
standard second quantization notation) \cite{SJD}:
\be\label{curop}
\hat{\vec j}(\vec r)=
\frac{ie}{2m}\left[\left(\vec\nabla\hat\psi^\dag\right)\hat\psi-\hat\psi^\dag\vec\nabla\hat\psi\right]-
\frac{e^2}{mc}\hat\psi^\dag\left(\vec A+\vec{\tilde A}\right)\hat\psi,
\ee
where together with an electromagnetic vector potential $\vec A$ (coupling to the charge $e$) a fictitious spin-dependent
vector potential $\vec{\tilde A}$ is introduced:
\be\label{fictiousA}
\vec{\tilde A}=-(\alpha mc/e)(-\hat\sigma^2,\hat\sigma^1,0),
\ee
where $c$ is the speed of light. On the other hand, the spin-current density operator ($z$-component of spin) remains the
same as in the case without spin-orbit interaction \cite{SJD,footnote3}:
\bea
\lefteqn{\hat{\vec j}^{s_z}(\vec r)=-\frac i{4m}\left[\hat\psi^\dag\hat\sigma^3\vec\nabla\hat\psi-
(\vec\nabla\hat\psi^\dag)\hat\sigma^3\hat\psi \right.}\qquad\qquad\qquad\qquad\qquad
\nonumber \\ & &
\left. -\frac{2ie}c\vec A\hat\psi^\dag(\vec r)\hat\sigma^3\hat\psi(\vec r)\right].
\label{sco}
\eea

\section{Spin-Hall current} \label{sec:sHc} %________________________________________________________________
Considering now a linear response regime for the spin current in the presence of an electric field $E_x$, we
derive a generalized Kubo-Greenwood formula for
the spin-Hall conductivity. In the Keldysh approach \cite{Keldysh} the spin current can be expressed as
\bea
\lefteqn{%\hspace{.07\textwidth}
\overline{\vec j^{s_z}(\omega)}=\label{curNonEq}} \\ && \nonumber
-\frac i{2m}\int_{-\infty}^\infty\frac{\ud E}{2\pi}
\overline{\Sp\left\{\left(\hat{\vec p}-\frac ec\vec A\right)\frac{\hat\sigma^3}2{\hat G}_{\mathrm K}
(E,E-\omega)\right\}},
\eea
where $\GK$ is the Keldysh component of the 2x2 matrix Green function;
$\Sp$ stands for the trace in both momentum and spin space.
Due to the Pauli matrix $\hat\sigma^3$ the diamagnetic term $\propto\vec A$ % in (\ref{sco}) 
gives a vanishing contribution\cite{footnote2} to the spin current in eq. (\ref{curNonEq}).
Then in first order perturbation in $\vec A$ we find
\bea
\lefteqn{\overline{\delta{\hat G}_{\mathrm K}(E,E-\omega)}=
\left(h_E-h_{E-\omega}\right)\label{KeldyshCond}}\qquad\qquad\nonumber\\ & & \times
\overline{
{\hat G}^E_{\mathrm R}\left[-\frac e{mc}\left(\hat{\vec p}-\frac ec\vec{\tilde A}\right)
\vec A(\omega)\right]{\hat G}^{E-\omega}_{\mathrm A}
},\quad
\eea
where $h_E=\tanh(E/2k_{\mathrm B}T)$, with $k_{\mathrm B}T\ll\EF$ the temperature and $\EF$ the Fermi
energy, and where ${\hat G}^E_{\mathrm{R/A}}=(E-\hat H+\EF\pm i0^+)^{-1}$
are the retarded and advanced Green functions, respectively.
From eqs. (\ref{curNonEq}) and (\ref{KeldyshCond}) we obtain a generalized Kubo-Greenwood formula for the
spin-Hall conductivity\cite{footnote1},
\be\label{shc}
\shs(\omega)=\frac e{2\pi m^2}
\overline{\Sp\left[\frac{\hat\sigma^3}2\hat p_y{\hat G}^E_{\mathrm R}\left(\hat p_x-\frac ec\tilde A_x\right)
{\hat G}^{E-\omega}_{\mathrm A}\right]},
\ee
where $E,\omega\ll\EF$.
In deriving (\ref{shc}) we have assumed that the electric field $E_x(\omega)=i\omega A_x(\omega)/c$ is applied along the
$x$-direction producing a perpendicular spin current along
$y$-direction, i.e.,
\be\label{spincurrent1}
\overline {j_y^{s_z}(\omega)}=\shs(\omega) E_x(\omega).
\ee
Next, after averaging over the random disorder, the spin-dependent
Green functions ${\hat G}_{\mathrm{R/A}}$ become diagonal in momentum representation \cite{Skvortsov},
$\overline{\langle\vec p|{\hat G}_{\mathrm{R/A}}^E|\vec p'\rangle}
=(2\pi)^2 \GRA^E(\vec p)\delta(\vec p-\vec p')$, with
\be
\GRA^E(\vec p)=\frac12\sum_{s=\pm1}\frac{\hat\sigma^0+s\hat M}{E-\xi(p)-s\alpha p\pm\frac i{2\tau}},\label{avg}
\ee
where $\hat M=(p_y\hat\sigma^1-p_x\hat\sigma^2)/p$
with ${\hat M}^2=\hat\sigma^0\equiv 1$, $\xi(p)=p^2/2m-\EF$.
{The expression (\ref{avg}) has been obtained in the self-consistent Born approximation.}
\begin{figure}[h]
\begin{center}
\psfrag{-evA}{$\hat{\vec j}_x$}\psfrag{j}{$\hat{\vec j}^{s_z}_y$}
\psfrag{p}{${\hat G}^E_{\mathrm R}$}\psfrag{p-q}{$G^{E-\omega}_{\mathrm A}$}
\includegraphics[width=.15\textwidth]{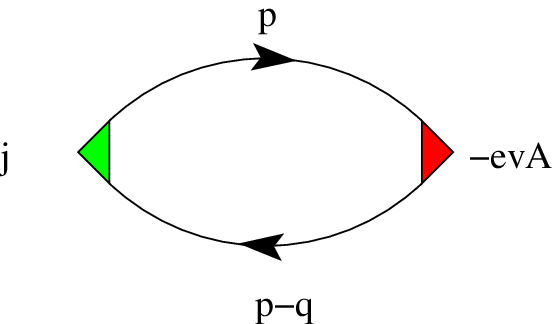}\quad
\includegraphics[width=.15\textwidth]{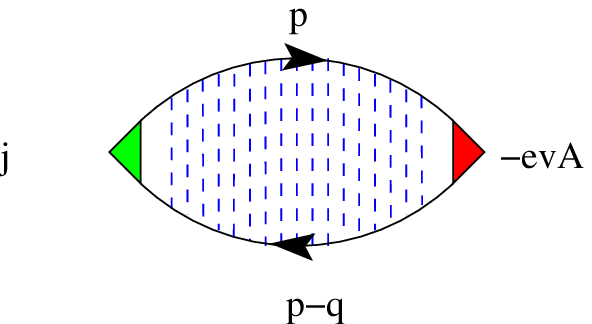}
\end{center}
\caption{Diagrams for the spin-Hall conductivity in the zero-loop approximation:
The right diagram represents the vertex correction.
Dashed lines denote averaging over the impurities.\label{fHallDiagram}}
\end{figure}

The diagrams of the standard perturbative approach \cite{AGD} can be generated
from (\ref{shc}) by expanding in the disorder potential $U$
and ``dressing'' the Green function lines with diffusons and Cooperons.
In this formalism \cite{AGD,Bergmann}, the resulting diagrams can be estimated
according to the number of loops, composed by Cooperon and diffuson lines (loop expansion).
{In the following sections we will expand $\shs$ in powers of $1/\pf l$,
where $l=\vf\tau$ denotes the mean free path, and evaluate  it up to  order $\frac{|e|}{(\pf l)^0}$ , that is,
we will neglect all terms with $n\ge1$ in the expansion
\be\label{series}
\shs=|e|\sum_n\frac{s_n}{(\pf l)^n},\quad\pf l\gg1.
\ee
Thus in our diagrammatic loop expansion we have to consider two classes of diagrams:
(i) diagrams with no loops (semiclassical approximation), and (ii) diagrams with one loop (weak localization corrections).
This is so because of the
non-standard situation arising from the spin dependence of the charge current vertex in (\ref{shc}),
which can be written as
\be\label{dcc}
\hat p_x-\frac ec\tilde A_x=\pf\hat n_x+\left(\hat p_x-\pf\hat n_x-\frac ec\tilde A_x\right),
\ee
where $\pf$ is the Fermi momentum, and $\hat{\vec n}={\hat{\vec p}}/p$ is an operator of the direction of momentum $\hat{\vec p}$.
The  contribution in the brackets on the right-hand side of (\ref{dcc}) is a correction of order $(\pf l)^{-1}\ll1$
to the main term $\pf\hat n_x$.
Thus, in  zero-loop approximation (see Sec.~\ref{sec:zeroLoop}) this correction leads to a contribution which is of
the same order  as the one coming from $\pf\hat n_x$ in the one-loop approximation (i.e. weak localization correction --
see Sec.~\ref{sec:WL}).
Hence, both of them are expected to give contributions of the same order  in $1/\pf l$ to the spin-Hall conductivity. 
The spin current vertex ${\hat\sigma}_z{\hat p}_{y}$, on the other hand, can always be approximated by
${\hat\sigma}_z\pf{\hat n}_y$ in the order considered in this work.

\section{Zero-loop approximation} \label{sec:zeroLoop} %___________________________________________________________
In zero-loop approximation we need to retain only the diagrams depicted in Fig.~\ref{fHallDiagram} with no loops.
%Both of them vanish in the limit $\alpha\to0$.
This approximation corresponds to the semiclassical (Boltzmann) limit.
To calculate the first diagram, we simply substitute ${\hat G}_{\mathrm{A/R}}$ in (\ref{shc}) by the averaged Green functions
(\ref{avg}). One can see that the main term of the charge current vertex (\ref{dcc}) does not contribute
in the first diagram. The remaining terms in the brackets in (\ref{dcc}) give
\be\label{sSHz}
\ssh0(\omega)=\frac{|e|}{8\pi}\times\frac{x^2}{x^2+\lambda^2},\quad\lambda=1-i\omega\tau ,
\ee
{where $x\equiv2\alpha\pf\tau$ is the dimensionless spin-orbit parameter which measures
the spin rotation between two consecutive impurity collisions.}
For $\omega=0$ this expression agrees with the qualitative result of Ref.~\cite{DJ} in the limit of small spin-orbit interaction
$m\alpha^2\tau\ll1$, as well as in the ballistic limit $\tau\to\infty$.

%\paragraph{Calculating the vertex correction diagram.}
Next we turn to the vertex correction, described by the second diagram in Fig.~\ref{fHallDiagram}. 
Again, it is crucial\cite{PrivateMSH}
to retain the entire correction of the charge current vertex (\ref{dcc}) (which amounts to include
curvature effects of the Fermi surface), since it will turn out now that this correction term  leads to an exact  cancellation of $\ssh0$ given in eq. (\ref{sSHz}).
We note that the
scattering off impurities is spin-independent, so that dashed lines do not alter the spin direction which simplifies the
calculation considerably.  For further evaluation it is convenient to apply the following identity for the Pauli matrices
\cite{Edelstein:93}
\be\label{borinoTozhdestvo}
\hat\sigma^0_{s_1s_2}\hat\sigma^0_{s_3s_4}=\frac12\sum_{\mu=0}^3\hat\sigma_{s_1s_3}^\mu\hat\sigma_{s_4s_2}^\mu
\ee
to every dashed line in the vertex diagram Fig.~\ref{fHallDiagram}, with $\alpha,\beta=0,\ldots,3$, and where, again,
$\hat\sigma^0$ is the unity matrix.  This trick reduces the vertex diagram to a geometric series consisting of terms
where the momentum and spin summation can be performed separately. These terms are given by
\be
\label{defXabDC}
X_{\mathrm D}^{\alpha\beta}=\frac1{2m\tau}
\Sp[\hat\sigma^\alpha G^E_{\mathrm R}(\vec p)\hat\sigma^\beta G^{E-\omega}_{\mathrm A}(\vec p)],
\ee
where from now on $\Sp$ stands for $\int\frac{\ud^2p}{(2\pi)^2} \Sp_{\mathrm{spin}}$.
%(the restriction
%to the Fermi surface can now be neglected, see below).
The off-diagonal elements of $X_{\mathrm D}^{\alpha\beta}$ vanish, while the diagonal ones are given by
\bea
\lefteqn{\nonumber X_{\mathrm D}^{00}=\frac1\lambda,\quad X_{\mathrm D}^{33}=\frac\lambda{x^2+\lambda^2},}\\ & &
\label{secondoX} X_{\mathrm D}^{11}=X_{\mathrm D}^{22}=\frac12\left[\frac1\lambda+\frac\lambda{x^2+\lambda^2}\right].
\eea
The expression for the vertex diagram can be split into three parts, and
the vertex diagram can be represented by two ``bubbles'' with a diffuson wavy line in between:
\bea
\lefteqn{\hspace{-.03\textwidth}
\psfrag{-evA}{$\hat{\vec j}_x$}\psfrag{j}{$\hat{\vec j}^{s_z}_y$}
\insfig{.33\textwidth}{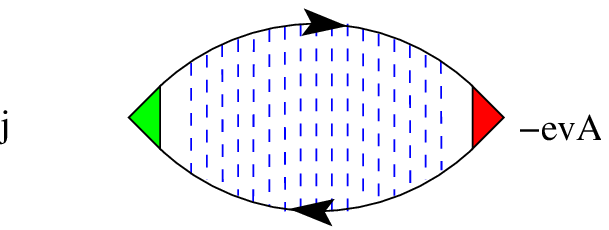}\hspace{-.22\textwidth}
=\sum_{\mu=0}^3%\hspace{-.02\textwidth}
\psfrag{GR}{}\psfrag{GA}{}\psfrag{lv}{}\psfrag{rv}{}\psfrag{Di}{$D^{\mu\mu}$}\psfrag{si}{$\hat\sigma^\mu$}
\insfig{.5\textwidth}{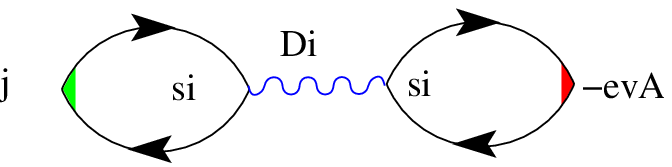}\hspace{-.265\textwidth}=\nonumber
}\\ & &\nonumber
\hspace{-.03\textwidth}=\frac e{2\pi m^2}\sum_{\mu=0}^3
\Sp\left[\frac{\hat\sigma^3p_y}2G^E_{\mathrm R}
(\vec p)\hat\sigma^\mu G^{E-\omega}_{\mathrm A}(\vec p)\right]\times\\ & &\times
D^{\mu\mu}\Sp\left[\hat\sigma^\mu G^E_{\mathrm R}(\vec p)\left(p_x-\frac ec\tilde A_x\right)G^{E-\omega}_{\mathrm A}(\vec p)\right],
\label{trucco}\eea
where $D^{\mu\mu}$ denotes the diffuson
%{I've inserted $\frac12$ in (\ref{cdRashba}) and eliminated $\frac12$ in (\ref{trucco})}
\be\label{cdRashba}
D^{\mu\mu}=\frac1{2m\tau}\frac1{1-X_{\mathrm D}^{\mu\mu}},
\ee
with $X_{\mathrm D}^{\mu\mu}$ given in (\ref{secondoX}).
Only the term with $\mu=2$ contributes to (\ref{trucco}). The two bubbles in (\ref{trucco}) are explicitly given by
\bea\label{lbub}
\Sp\left[\frac{\hat\sigma^3p_y}2G^E_{\mathrm R}(\vec p)\hat\sigma^2G^{E-\omega}_{\mathrm A}(\vec p)\right]=
%\frac{\pi\nu x\tau\pf}{x^2+\lambda^2},
\frac{xm\tau\pf/2}{x^2+\lambda^2},
\eea
and% \cite{footnote4}
\bea
\Sp\left[\hat\sigma^2G^E_{\mathrm R}(\vec p)\left(p_x-\frac ec\tilde A_x\right)G^{E-\omega}_{\mathrm A}(\vec p)\right]=
\nonumber\\
=\frac{\alpha m^2\tau}\lambda\frac{x^2}{x^2+\lambda^2}.
\label{rbub}
\eea
Inserting these expressions into (\ref{trucco}), we finally obtain for the vertex correction
of the spin-Hall conductivity
\be\label{vdov}
\ssh1(\omega)=-\frac{|e|}{8\pi}\frac{x^2}{x^2+\lambda^2}
\frac{x^2/2}{\left(1-2i\omega\tau\right)\frac{x^2}2-i\omega\tau\lambda^2}.
\ee
Adding now eqs. (\ref{sSHz}) and (\ref {vdov}) we obtain 
the spin-Hall conductivity in zero-loop approximation,
\bea % checked for the correspondence with (\ref{sigmaRu})=(2.30) from the notes.
\lefteqn{\shs(\omega)=\ssh0(\omega)+\ssh1(\omega)= \nonumber}\\ & & \label{spinHallconductivity}
=\frac{|e|}{8\pi}\frac{x^2}{x^2+\lambda^2}\left\{1-\frac{x^2/2}{(1-2i\omega\tau)x^2/2-i\omega\tau\lambda^2}\right\},\quad
\eea
which is the same result as obtained in Ref.~\cite{MSH} (see eq. (25) therein) by the equation
of motion approach.
Thus, in  zero-loop approximation the two diagrams
in Fig.~\ref{fHallDiagram} cancel each other, so that the leading contribution to the spin-Hall conductivity vanishes in
the limit $\omega=0$, in agreement with \cite{ibmNew,MSH,Olya,Sasha,RashbaSaysZero,RS0}.
A similar cancelation has been observed also for the case of the electric conductivity \cite{Schwab}.
In both these cases  the ``anomalous'' term $\propto\tilde A$ in the charge current operator (\ref{curop}) is cancelled by
the vertex correction at $\omega=0$.
% (and for zero magnetic field).
Thus,
the second diagram in Fig.~\ref{fHallDiagram} can be interpreted as a renormalization of the charge
current vertex which
results in the cancellation of the ``anomalous'' term
$\propto\tilde A_x$.

\section{Weak localization correction} \label{sec:WL} %______________________________________________________________
At the end of Sec.~\ref{sec:sHc} we argued that in order to reach a given accuracy, we have to include the
contribution of one-loop diagrams in our calculation of $\shs$.
In this section we demonstrate this now by calculating the weak localization correction to $\shs$ explicitly.

Similar to  the case of the charge conductivity\cite{Edelstein:94},
the weak localization correction to the spin-Hall conductivity is given by three diagrams depicted in Fig.~\ref{fWL}.
For simplicity, we calculate them at zero frequency $\omega=0$ only;
to simplify notations we omit the energy superscript of Green functions, assuming
$\GRA\equiv G^E_{\mathrm{R/A}}$ below.

\begin{figure}
\begin{center}
\psfrag{-evA}{$\hat{\vec j}_x$}\psfrag{j}{$\hat{\vec j}^{s_z}_y$}
\psfrag{p}{}\psfrag{p-q}{}\psfrag{q-p}{}
\insfigh{2cm}{2cm}{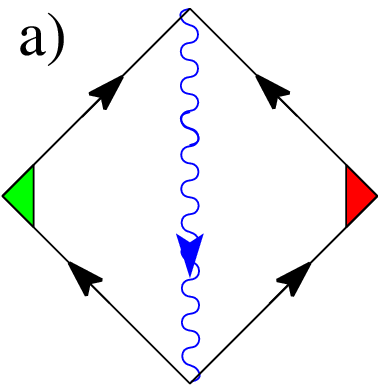}\quad
\insfigh{2cm}{2cm}{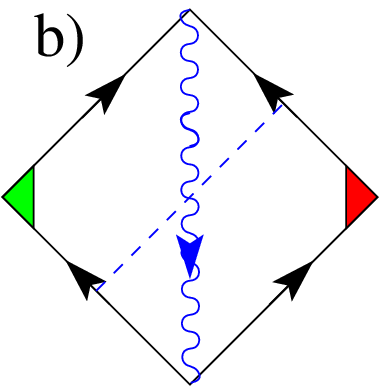}\quad
\insfigh{2cm}{2cm}{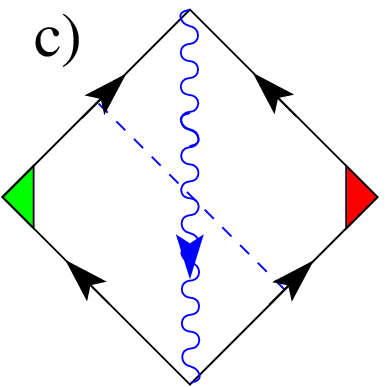}
\end{center}
\caption{Weak localization diagrams. The wavy line represents the Cooperon (\ref{ourC}).
Each diagram contains one spin current vertex (left) and one charge current vertex (right).
%Spin and charge current vertices are renormalized according to eq.~(\ref{rle}) and (\ref{rri}).
\label{fWL}}
\end{figure}

Using (\ref{borinoTozhdestvo}), we may separate the spin indices of a Hikami box from those  of the Cooperon, like we did
above in (\ref{trucco}). Then the total contribution of the three diagrams in Fig.~\ref{fWL} can be written as
\be
\ssh2(\omega)=\frac e{2\pi m^2}\sum_{\gamma,\gamma'=0}^3\iqd
C^{\gamma\gamma'}(\vec q)V^{\gamma\gamma'}(\vec q),\label{ivdsl}
\ee
where $C^{\gamma\gamma'}(\vec q)$ are the matrix elements of the Cooperon defined as
\be\label{ourC}
C^{\gamma\gamma'}(\vec q)=\frac1{2m\tau}\left[\frac{X_{\mathrm C}(\vec q)}{1-X_{\mathrm C}(\vec q)}\right]_{\gamma\gamma'},
\gamma,\gamma'=0\ldots3,
\ee
\be
X_{\mathrm C}^{\alpha\beta}(\vec q)=\frac1{2m\tau}
\Sp[\hat\sigma^\alpha G_{\mathrm R}(\vec p)\hat\sigma^\beta(G_{\mathrm A}(\vec q-\vec p))^T].\quad
\ee
In eq.~(\ref{ivdsl}), $V^{\gamma\gamma'}(\vec q)$ denotes the  Hikami box:
\be\label{hbgen}
V^{\gamma\gamma'}(\vec q)=V_a^{\gamma\gamma'}(\vec q)+V_b^{\gamma\gamma'}(\vec q)+V_c^{\gamma\gamma'}(\vec q).
\ee
Each of the three terms in eq.~(\ref{hbgen}) corresponds to one diagram in Fig.~\ref{fWL} plus the corresponding
diagram in Fig.~\ref{fWLren} (see eq.~(\ref{rle}) below).

In the rest of this section we assume $x^2\ll1$. This permits us to  approximate (\ref{ivdsl}) by
\be\label{wlctc}
\ssh2(\omega)=\frac e{2\pi m^2}\sum_{\gamma=0}^3V^{\gamma\gamma}(0)
\int_{0<q<1/l}\frac{\ud^2q}{(2\pi)^2} C^{\gamma\gamma}(\vec q).
\ee
Here we have neglected the off-diagonal elements of the Cooperon, as well as
 the $q\ne0$  corrections to the 
Hikami box in (\ref{hbgen}). This is justified since those 
terms do not contain 
logarithms in $x$ (or dephasing length),
in contrast to the diagonal Cooperon contributions in (\ref{wlctc}) [see eqs. (\ref{iC}),(\ref{logs}) below].

For $q=0$, the three contributions to the Hikami box in eq.~(\ref{hbgen}) become
\be\label{WLa}
V_a^{\gamma\gamma}=\Sp\left\{
L(\vec p)\hat\sigma^\gamma\left[\hat\sigma^{\gamma}R(-\vec p)\right]^T\right\},
\ee
\be
\label{WLbc}
V_{b(c)}^{\gamma\gamma}=\frac1{2m\tau}\sum_{\mu=0}^3A_{b(c)}^{\gamma\mu}B_{b(c)}^{\mu\gamma},
\ee
where for the left (L) and right (R) vertex parts we use the notations
\bea
\lefteqn{L(\vec p)=\GA(\vec p)p_y\frac{\hat\sigma^3}2\GR(\vec p),}\\ & &
\begin{split}
R(-\vec p)&=\GR(-\vec p)\left(-p_x-\frac ec\tilde A_x\right)\GA(-\vec p)\approx\label{unapr}\\
&\approx\GR(-\vec p)\left(-n_x\pf\right)\GA(-\vec p).\label{sprava}
\end{split}\qquad
\eea
Retaining only leading order, we have approximated the charge current vertex in (\ref{sprava}) in accordance with the
discussion at the end of Sec.~\ref{sec:sHc}. [The same approximation is made in the calculation of the weak localization
correction to the charge conductivity\cite{WLcorr79}.] The quantities $A$ and $B$ in (\ref{WLbc}) are defined as
\bea
&A_b^{\gamma\mu}=&\Sp\left\{L(\vec p)\hat\sigma^\gamma\GA^T(-\vec p)\hat\sigma^\mu\right\},\label{wldba}\\ \label{wldbb}
&B_b^{\mu\gamma}=&\Sp\left\{\hat\sigma^{\gamma}R(-\vec p)\left[\GA(\vec p)\hat\sigma^\mu\right]^T\right\},  \\
&A_c^{\gamma\mu}=&\Sp\left\{ L(\vec p)\hat\sigma^\mu\left[\hat\sigma^\gamma\GR(-\vec p)\right]^T\right\},  \\
&B_c^{\mu\gamma}=&\Sp\left\{\hat\sigma^\mu\GR(\vec p)\hat\sigma^{\gamma}\left[R(-\vec p)\right]^T\right\}.
\eea
Due to the symmetry properties
\be\label{sopr}
\begin{split}
(A_b^{2\mu})^\ast=&-A_c^{2\mu},\quad (B_b^{\mu2})^\ast=-B_c^{\mu2},\\
(A_b^{\gamma\mu})^\ast=&A_c^{\gamma\mu},\quad (B_b^{\mu\gamma})^\ast=B_c^{\mu\gamma},\quad\gamma\ne2,
\end{split}
\ee
we see that for all $\gamma\ V^{\gamma\gamma}_b=(V^{\gamma\gamma}_c)^\ast$ in (\ref{WLbc}).

In addition, for every diagram in Fig.~\ref{fWL} one needs to consider the corresponding
renormalization of the spin current vertices,
which turns out to be of the same order in $(\pf l)^{-1}$ as the bare $ L(\vec p)$,
\be\label{rle}
\begin{split}
\tilde L(\vec p)=L(\vec p)+&\sum_{\mu=0}^3\GA(\vec p)\hat\sigma^\mu\GR(\vec p)\times\\ 
\times&D^{\mu\mu}\Sp\left\{\hat\sigma^\mu L(\vec p)\right\}.
\end{split}
\ee
The corresponding diagrams are depicted in Fig.~\ref{fWLren}.
Their contributions add to the ones of the bare diagrams in Fig.~\ref{fWL}.
Note that the symmetry properties (\ref{sopr}) remain valid when the renormalization (\ref{rle}) is taken into account in
the expressions for $A$ and $B$.

Similary, the charge current vertex ($\propto R(\vec p)$) gets renormalized, but unlike before this renormalization is of higher order in
$(\pf l)^{-1}$ [see at the end of Sec.~\ref{sec:zeroLoop}], and hence can be neglected. 
Still, we note that like  in Sec.~\ref{sec:zeroLoop}
one can demonstrate that  this
renormalization of $R(\vec p)$
results in cancellation of the ``anomalous'' term $\propto\tilde {A_x}$ in the charge current vertex  in (\ref{unapr}). 
%This ``anomalous'' term has been anyway ignored in (\ref{sprava}).

\begin{figure}
\begin{center}
\includegraphics[width=\columnwidth]{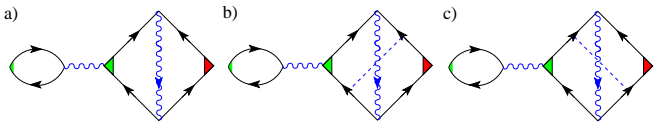}
\end{center}
\caption{The weak localization diagrams from Fig.~\ref{fWL} with the renormalized spin current vertex 
($\propto L(\vec p)$), see eq.~(\ref{rle}).
The analogous renormalization of the current vertex ($\propto R(\vec p)$) is of higher order in 
$(\pf l)^{-1}$ and can be neglected.
\label{fWLren}}
\end{figure}

Now we have to obtain an expression for $\iqd C^{\gamma\gamma}(\vec q)$ in (\ref{wlctc}).  For this we can make use of
the results of Ref.~\cite{Skvortsov}, where an analogous approach has been used to study the weak localization correction
to the charge conductivity in the same model system.  The Cooperon derived in Ref.~\cite{Skvortsov} is connected with $C$
in eq. (\ref{ourC}) via
\be\label{toSkvor}
C^{\gamma\gamma'}(\vec q)=\frac14\sum_{\alpha,\beta,\mu,\lambda=1}^2C^{\alpha\lambda}_{\beta\mu}
\hat\sigma^\gamma_{\beta\alpha}\hat\sigma^{\gamma'}_{\lambda\mu},
\ee
where $C^{\alpha\lambda}_{\beta\mu}$ is the Cooperon from Ref.~\cite{Skvortsov}.
Then, from eqs.~(13) and (14) in Ref.~\cite{Skvortsov}
we obtain that  
\be\label{iC}
\iqd C^{\gamma\gamma}(\vec q)=\frac2{\pi l^2m\tau}(f,f,\log\frac{L_\phi}l,f),
\ee
where \cite{Skvortsov} $L_\varphi$ stands for the dephasing length, and
\be\label{logs}
f=\begin{cases}
    \ln\frac{L_\varphi}l , & \mathrm{for}\quad x\ll\frac l{L_\varphi}, \\
    \ln\frac1x , & \mathrm{for}\quad\frac l{L_\varphi}\ll x\ll1.
\end{cases}
\ee
We note that due to this logarithmic dependence on $x$ the Cooperon itself
can become large for sufficiently small $x$ and large dephasing length ${L_\phi}$. However,
this has no consequences for the spin Hall conductivity eq. (\ref{wlctc}),
since the Hikami box contributions add up to zero, as we shall see next.
Indeed, after some lengthy but straightforward calculation we find for
the diagonal terms of the Hikami box \cite{spinRenDNC}
\be
V_a^{\gamma\gamma}=(0,0,0,0)+O(x^4),
\ee
\bea
\begin{split}
V_b^{\gamma\gamma}&=-V_c^{\gamma\gamma} \\
&=-\frac{im\pf^2\tau^3}{16}(x^2,8-15x^2,-x^2,8-11x^2).
\end{split}
\eea
Quite remarkably, the terms independent of $x$ come from the vertex renormalization (2nd term
on the rhs of  eq. (\ref{rle})), where the only non-zero contribution arises
from the diffuson matrix element $D^{{22}}\propto 1/x^{2}$.
Each of the terms $V_b$ and $V_c$, when inserted into eq. (\ref{wlctc}), gives a contribution to $\ssh2$ 
$\propto e$, which corresponds to the term $n=0$ in (\ref{series}). Adding up all  three terms of the Hikami box
we obtain
$V^{\gamma\gamma}(0)=0$ in (\ref{hbgen}). Thus, we see that the weak
localization contribution to the spin Hall conductivity, i.e.
$\ssh2$ (see eq. (\ref{wlctc})),  vanishes in the limit $\omega=0$. 

The evaluation of the zero- and one-loop diagrams performed in this and the preceding sections is valid up to the order
$1/\pf l$. If one would like to go beyond that order (which is not attempted here), many more corrections (e.g. of the
type $1/\EF\tau$) and many more diagrams had to be taken into account. For instance, one would also need to go beyond the
self-consistent Born approximation for the self-energy and to retain crossing diagrams.  Of course, the weak localization
contribution which includes the Cooperon (resulting from time-reversed paths) is special in the sense that (apart form
the log-dependence) it is sensitive to phase coherent effects probed by an Aharonov-Bohm flux.  Thus, other
contributions, even if they are of the same order in $1/\pf l$, could be ignored provided one would be interested in the
phase sensitive contributions only.

\section{Spin current and spin precession} \label{sec:mustDie} %________________________________________________
%\msc{correct also abstract, introduction, and conclusion}
We have seen that in perturbation theory the spin-Hall conductivity vanishes in the zero frequency limit, in leading
($\propto(1/\pf l)^0$) {\it and} subleading ($\propto(1/\pf l)^1$) order.
This result suggests that the vanishing of the spin current is an exact property of the system under consideration.

Indeed, we give now a simple argument to support this claim, based on the generalization of the
Hamiltonian (\ref{hamrash}):
\be\label{CH}\begin{split}
\hat H'(\hat{\vec p},\vec r)=\frac{\hat p^2}{2m}+&
U(\vec r)+\alpha\left(\hat\sigma_1\hat p_y-\hat\sigma_2\hat p_x\right)+\\
+& \beta\left(\hat\sigma_1\hat p_x-\hat\sigma_2\hat p_y\right)+e\vec r\vec E,
\end{split}\ee
where $\vec E$ is the applied static electric field, and
$\beta$ is the amplitude of Dresselhaus spin-orbit interaction.

From the Heisenberg equation of motion for the spin $\hat{\vec s}=\hat{\pmb\sigma}/2$ of
the electron, i.e.  $\dot{\hat{\vec s}}(t)\equiv\frac\ud{\ud t}{\hat{\vec s}}(t)=i[\hat H',\hat{\vec s}](t)$, we obtain a
relation between spin precession and spin current \cite{footnoteSJD}.
\be\label{emsx}\begin{split}
-\frac1{2m}\dot{\hat s}_x(t)=\alpha\hat j_x^{s_z}(t)+\beta\hat j_y^{s_z}(t),\\
-\frac1{2m}\dot{\hat s}_y(t)=\alpha\hat j_y^{s_z}(t)+\beta\hat j_x^{s_z}(t),
\end{split}\ee
or
\be\label{emsxII}\begin{split}
\hat j_x=-\frac1{2m}\frac{\alpha\dot{\hat s}_x-\beta\dot{\hat s}_y}{\alpha^2-\beta^2},\\
\hat j_y=-\frac1{2m}\frac{\beta\dot{\hat s}_x-\alpha\dot{\hat s}_y}{\beta^2-\alpha^2},\end{split}\quad\alpha\ne\beta.
\ee
% which is valid also for disordered systems, and, moreover, in the presence of a constant external electric field
% \cite{footnoteSJD}.  In this case we have $\hat H'=\hat H+e\hat xE_x$, where an electric field $E_x$ is applied along the
% $x$ direction, as considered in the previous section (zero-frequency limit), and $\hat H$ is given in
% eq. (\ref{hamrash}).
%
The uniform spin
current operator ($z$-component of the spin) is obtained from (\ref{sco}) and given in first quantization notation by
%$\hat{\vec j}^{s_z}=(1/2)\{\hat{\vec v},\hat s_z\}=\hat{\vec p}\hat s_z$,
$\hat{\vec j}^{s_z}=(1/2)\{\hat{\vec v},\hat s_z\}$,
with (spin-dependent) velocity operator $\hat{\vec v}=i[\hat H',\vec r]$.

From the exact relations (\ref{emsxII}) we see that
the spin current can be experimentally accessed by measuring the spin precession or the change of magnetization in time,
which can be done e.g. with optical methods \cite{kato:0403407,ganichev:0403641,AwschScience04}.

In case of $\alpha=0$ or $\beta=0$  (\ref{emsxII}) provides a simple
physical interpretation of the spin current:
The spatial $x$ ($y$)-component of the spin current which carries the $z$-component of the spin
is, up to a coupling constant, given by the total time derivative of the $x$ ($y$)-component of the spin.

Next, let us consider the expectation value (including disorder average) of (\ref{emsxII}) for $t\to\infty$.
%$\langle{\dot{\hat s}}_{k}\rangle(t)=-2m\alpha\langle\hat j_k^{s_z}\rangle(t),\quad t\to\infty.$
The presence of  disorder is expected to provide a relaxation mechanism such that the system can reach a steady nonequilibrium state (after some
transients) when driven by an external electric field (assumed to be constant in time). The weaker the disorder the
longer it takes to reach this stationary limit, but for any finite amount of disorder (finite density of random impurities), it will be reached eventually.
In particular, the magnetization $\langle\hat{\vec s}\rangle(t)$ of such a state also 
reaches a stationary value (possibly different from zero) and does not depend on time anymore. [Phenomenologically,
the approach to such a state is
described by a Bloch equation for the spin dynamics; see also below.]
Thus, it immediately follows that the rate of
magnetization change, $\langle \dot{\hat{\vec s}}\rangle(t)$, must vanish in such a steady state, and, consequently,
the spin current vanishes as well.
In other words, under the stated conditions
% {\it the uniform spin current is non-zero  if and only if the  magnetization
%   does not reach a stationary state for $t \to\infty$}.
\emph{the uniform spin current vanishes in the long time limit if and only if the
magnetization reaches a stationary state,} i.e.
\be\label{theorem}
\lim_{t\to\infty}\langle\hat{\vec j}^{s_z}\rangle(t)=0\Longleftrightarrow
\lim_{t\to\infty}\langle\hat s_k\rangle(t)=\mathrm{const.},\quad k=x,y.
\ee
The result (\ref{theorem}) holds for any form of the static (spin-independent) impurity potential $U(\vec r)$
in the Hamiltonian (\ref{CH}), including
the special case of isotropic scattering potentials considered in previous sections.
%It also remains valid if one takes into account Coulomb interaction between electrons.
Also, the argument is valid for
any strength of the electric field, and, hence, applies to the special case of linear response considered in the previous
section [with
$\langle\hat j_y^{s_z}\rangle\equiv \overline {j_y^{s_z}}$]. Thus, from $0=\langle \hat j_y^{s_z}\rangle(t\to\infty)= \shs(\omega=0)E_{x}$ we conclude that the spin Hall
conductivity $\shs(\omega=0)$ vanishes exactly (i.e. in all orders of the disorder potential) for any finite amount of
disorder.

In the special case\cite{BF} $\alpha=\beta$,
% (\ref{emsx}) reduces to
% \be-\frac{\dot{\hat s}_x(t)}{2m\alpha}=-\frac{\dot{\hat s}_y(t)}{2m\alpha}=\hat j_x^{s_z}(t)+\hat j_y^{s_z}(t),\ee
% so that we cannot deduce whether the spin-Hall current is zero or not in the limit of $t\to\infty$.
it is convenient to rewrite $\hat H'$ in (\ref{CH}) in the coordinate system, rotated counterclockwise by $\pi/4$ in the
$xy$-plane:
\be\label{RDrotH}
\begin{split}
\hat H_{\mathrm R}(\hat{\vec p}',\vec r')\equiv
\hat H'(R_{\pi/4}\hat{\vec p},R_{\pi/4}\vec r)&=\\
=\frac{\hat p^{'2}}{2m}-2\alpha\hat\sigma_2'\hat p_x'+U'(\vec r'\,)+e\vec r'\vec E'\equiv
\hat H_{\mathrm R0}+&e\vec r'\vec E',
\end{split}
\ee
where
\be
R_{\pi/4}=\left(\begin{array}{ccc}
 \frac1{\sqrt2}&\frac1{\sqrt2}&0\\
-\frac1{\sqrt2}&\frac1{\sqrt2}&0\\
0&0&1\end{array}\right),
\ee
and for convenience we have introduced an alternative set of Pauli matrices
\be\hat\sigma_{12}'=\frac1{\sqrt2}(\hat\sigma_2\pm\hat\sigma_1),\quad{\hat\sigma}_3'\equiv\hat\sigma_3.\ee

Suppose before switching on the electric field (at $t=0$) the system was described by an equilibrium density matrix
$\hat\rho_0=e^{-{\hat H}_{\mathrm R0}/T}/Z$. [It implies $\langle\hat{\vec j}^{'s_z}\rangle(t=0)=0$.]
Then the spin current in the rotated system is given by
\be\label{DLsays}
\langle\hat{\vec j}^{'s_z}\rangle(t)=\Sp\left[\frac{\hat\sigma_3}2
\frac{\hat{\vec p}}me^{-i\hat H_{\mathrm R}t}\hat\rho_0e^{i\hat H_{\mathrm R}t}\right].
\ee
Since 
\be\begin{split}
\Sp_\mathrm{spin}[\hat\sigma_k'\hat H_{\mathrm R}]=0,\quad\Sp_\mathrm{spin}[\hat\sigma_k'\hat H_{\mathrm R0}]=0,\quad k=1,3,\\
\Longrightarrow\Sp_\mathrm{spin}\left[\frac{\hat\sigma_3}2\frac{\hat{\vec p}}me^{-i\hat H_{\mathrm R}t}\hat\rho_0e^{i\hat H_{\mathrm R}t}\right]=0,\quad\forall t,
\end{split}\ee
so that $\langle\hat{\vec j}^{s_z}\rangle(t)=0$  for all $t>0$. Thus, once $\langle\hat{\vec j}^{s_z}\rangle$ is zero at
$t=0$, it remains zero forever due to the fact that $({\hat s}_y-{\hat s}_x)$ is a conserved quantity for the
Hamiltonian (\ref{CH}).% (\ref{RDrotH}).

We consider now $\beta=0$.
The approach of $\langle \dot{\hat{\vec s}}\rangle(t)$ to its stationary value (in the linear response
regime) can be easily illustrated by
taking the inverse Laplace transform of eqs. (\ref{spincurrent1}) and (\ref{spinHallconductivity}) (restricting ourselves to the Boltzmann value). This involves solving a cubic equation for obtaining the poles with respect to $\omega$.
The resulting expression for $\langle\dot{{\hat s}}_y\rangle(t)$  consists of two parts (too lengthy
to be written down here), one coming from the real pole and an oscillatory one coming from
the pair of complex conjugate poles.
For $x^{2}\ll1$ and $t\gg\tau$, only the first part is relevant, which has no oscillations and decays exponentially to zero:
\be\label{decay}
\langle\dot{{\hat s}}_y\rangle(t)=-2m\alpha \langle\hat j_y^{s_z}\rangle(t)=
-\frac{|e|}{4\pi}m\alpha  x^2  E_x \, e^{-t/T},\quad t\gg\tau,
\ee
where $T/2=(\Delta^2\tau)^{-1}$ is the well-known Dyakonov-Perel spin relaxation time \cite{DyakonovPerel}.
For $x^2\gtrsim1$ and $t\lesssim\tau$, the oscillatory part in $\langle\dot{\hat s}_y\rangle(t)$
becomes dominant, with period $1/\Delta$
and exponential decay with rate $1/\tau$; its time dependence for a particular value of $x$ is illustrated in Fig.~\ref{fDotSdecay}.
\begin{figure}[h]
\psfrag{xlabel}{$t/\tau$}\psfrag{ylabel}{$-\langle{\dot {\hat s}}_y(t)\rangle\frac{4\pi}{|e|m\alpha E_x}$}
\includegraphics[width=\columnwidth]{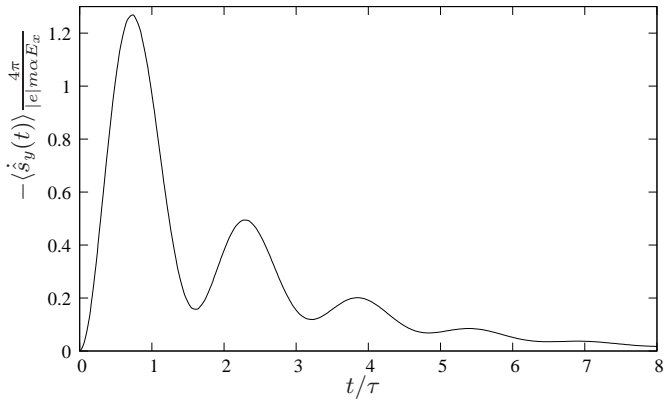}
\caption{The time evolution of $\langle{\dot {\hat s}}_y(t)\rangle$ for $x=4$.
The period of oscillation is $1/\Delta$, and the exponential decay time is $\tau$.
Note that for any $x$, $\langle{\dot {\hat s}}_y(t)\rangle=0$ at $t=0$. \label{fDotSdecay}}
\end{figure}

The above consideration can be  generalized to the case with electron-electron interaction.
In this case,  the total spin of the system, $\sum_i{\hat s}_{i,k}$, and the  total spin current,
$\sum_i{\hat j}_{i,k}^{s_z}$,  $k=x,y$, obey the same equation as before, i.e. eq. (\ref{decay}). Thus, the
same
argument goes through  as well, showing that the spin current must vanish
in the stationary limit also for interacting systems--provided this limit exists, which, again, we expect to be the case in the presence of any finite amount of disorder.

Finally, in the absence of disorder, no stationary limit of $\langle {\hat{\vec s}}\rangle(t)$ can be reached, i.e.  the
magnetization can change indefinitely, and thus there can be a finite spin current in this very particular case. [We note
that the physically observable quantity, the magnetization change, vanishes for vanishing Rashba coupling.]
However, when the spin current approaches a non-zero but constant value for $t\to\infty$, as obtained
in the linear response regime for a clean system \cite{culcer:0309475},
the magnetization $\langle {\hat s_k}\rangle(t)$ actually grows linearly in time in the asymptotic regime
$t\to\infty$. This, of course, shows a breakdown of the linear response approximation in this case since the bound for a
spin 1/2, i.e. $|\langle {\hat s}_{k}\rangle(t)|\leq1/2$, is violated. The terms beyond linear response would restore the
bounded and oscillatory behavior of $\langle {\hat s}_{k}\rangle(t)$.

\section{Conclusions}
In conclusion, we have calculated the Boltzmann and weak localization contributions to
the spin-Hall conductivity $\shs$ in an infinite-size 2D disordered system with Rashba spin-orbit
interaction. 
{In our calculation
we had to consider the weak localization contribution to the $\shs$, together with the Boltzmann contribution,
since diagrams belonging to these two contributions have the same order of magnitude.}
In the diffusive and static limit, the spin-Hall conductivity vanishes. 
%% We have shown this
%% by explicit diagrammatic evaluation of the Kubo formula for the spin Hall conductivity
%%  in perturbation theory for the Boltzmann value and the weak localization {correction}.
We have shown that the spin current is given by the time derivative of the magnetization change and have given general
arguments why the spin current must vanish for any finite amount of disorder that permits to reach a stationary limit of
the system when driven by a constant electric field. It will be interesting to see how far our arguments can be extended
to other types of spin orbit interaction.

%%%%%%%%%%%%%%%%%%%%%%%%%%%%%%%%%%%%%%%%%%%%%%%%%%%%%%%%%%%%%%%%%%%%%%%%%%%%%%%%%%%%%%%%%%%%%%%%%%%%%%%%%%%%%%%%%%%%%%
\section {Acknowledgments}
{We thank B.~Altshuler, S.~Erlingsson, J.~Schliemann, M.~Skvortsov, V.~Yudson, V.~Golovach, and in particular
E.~Mishchenko and A.~Shytov for helpful discussions. This work is supported by the Swiss NF, NCCR Nanoscience Basel, EU
Spintronics, US DARPA, and ONR.}
\bibliography{spinHall,refs.aps,spinCurrent,spintronics,spintronicsExp,spintronicsTheo}

\begin{thebibliography}{47}
\bibitem[{foo({\natexlab{a}})}]{footnote3}
\bibinfo{note}{For a discussion of the problems related with a proper
  definition of the spin current (which is not a conserved quantity) see also
  \cite{rashba03:241315R,rashba:0404723}.}
\bibitem[{foo({\natexlab{b}})}]{footnote2}
\bibinfo{note}{We note that the spin current (\ref{curNonEq}) contains no
  contribution $\propto(\GR-\GA)$, which for the charge current would be
  present and cancel the non-vanishing diamagnetic term.}
\bibitem[{foo({\natexlab{c}})}]{footnote1}
\bibinfo{note}{Eq. (\ref{shc}) can also be derived directly (without Keldysh
  formalism), by starting from the standard form of the Kubo formula for the
  conductivity, see e.g. Eq. (1) in \cite{DJ}.}
\bibitem[{spi()}]{spinRenDNC}
\bibinfo{note}{It turns out that the vertex renormalization of the spin
  current, although being non-zero, does not contribute to the diagonal
  elements of $V_{a,b,c}$.}
\bibitem[{foo({\natexlab{d}})}]{footnoteSJD}
\bibinfo{note}{We note that relation (\ref{emsx}) is the uniform limit of eqs.
  (31) and (32) in Ref.~\cite{SJD}, which form the basis of the conclusions
  reached here.}
\bibitem[{BF()}]{BF}
\bibinfo{note}{The case when $\alpha=\beta$ is discussed in
  \cite{schliemann03:146801} and in \cite{schliemann03:165311}.}
\end{thebibliography}
\end{document}